\documentclass[a4paper]{jpconf}
\usepackage{graphicx}
\begin{document}
\title{Torsional oscillators and the entropy dilemma of putative supersolid $^4$He}

\author{M J Graf$^1$, A V Balatsky$^1$, Z Nussinov$^2$, I Grigorenko$^1$, S A Trugman$^1$}
\address{$^1$Theoretical Division, Los Alamos National Laboratory, Los Alamos, New Mexico 87545, USA}
\address{$^2$Department of Physics, Washington University, St.~Louis, Missouri 63160, USA}
\address{LA-UR-08-04023}

\ead{graf@lanl.gov}

\begin{abstract}
Solid $^4$He is viewed as a nearly perfect Debye solid. Yet, recent calorimetry measurements by the PSU group 
(J.~Low~Temp.~Phys.~{\bf 138}, 853 (2005) and Nature~{\bf 449}, 1025 (2007))
indicate that at low temperatures the specific heat has both cubic and linear contributions.
These features appear in the same temperature range where measurements of the torsional oscillator 
period suggest a supersolid transition.
We analyze the specific heat and compare the measured with the estimated 
entropy for a proposed supersolid transition with 1\% superfluid fraction and find that the observed entropy
is too small. We suggest that the low-temperature linear term in the specific heat is due to a glassy state
that develops at low temperatures and is caused by a distribution of tunneling systems in the crystal.
We propose that dislocation related defects produce those tunneling systems. Further, we argue 
that the reported putative mass decoupling, that means an increase
in the oscillator frequency, is consistent with a glass-like transition. The glass scenario offers an
alternative interpretation of the torsional oscillator experiments in contrast to the supersolid scenario
of nonclassical rotational inertia.
\end{abstract}

\section{Introduction}
The discovery of an anomalous signature in torsional oscillator measurements of solid $^4$He has re-ignited the search 
for the enigmatic quantum state of supersolidity \cite{Kim_Chan}. 
Several groups using torsional oscillators report a drop in the resonance period around 0.15 K, 
with details depending on the amount of $^3$He impurity concentration present in $^4$He \cite{Kim_Chan,Rittner2006,Rittner2007,Kondo2007,Aoki2007}. 
The observed signal shows hysteresis effects depending on cooling and warming procedures or rim velocities, 
as well as annealing dependence. At this point it is far from clear if the observed phenomenon is due to the onset 
of supersolidity or the interplay of a dislocation glass or both.
Very recent Monte Carlo simulations suggest that a quasi-one-dimensional supersolid can occur along the axis of 
screw dislocations \cite{Boninsegni2007}. 
However, it is difficult to imagine how such a network of one-dimensional supersolid phases is the origin for reports 
of as much as $\sim 20$\% of supersolid fraction as seen in recent torsional oscillator experiments \cite{Rittner2008}. 
In this work, irrespective of the supersolid or glassy origin of the observed anomalies, 
we assume that only a small fraction of the perfect crystal undergoes a phase transition.

So far x-ray and neutron scattering experiments have failed to observe any evidence of 
a supersolid condensate fraction or changes in the lattice parameter, Debye-Waller factor, or phonon dispersions 
between 55 mK and 500 mK, where the transition is reported to occur \cite{Rittner2007,Adams2007,Diallo2007,Blackburn2007,Blackburn2008}. 
On the other side, specific heat \cite{Lin2008} 
and elastic shear modulus \cite{Day2007} 
measurements exhibit anomalies in the same temperature
range as torsional oscillators.

Here we discuss how a small fraction of the sample, undergoing a
glassy transition around 0.12 K, can account for the observed linear
$T$ dependence in the specific heat of an otherwise perfect Debye
solid. We emphasize that the physical picture invoked is not
that of a conventional glass, where the entire sample undergoes a
glass transition. Rather a small subsystem of the crystal, e.g.,
dislocations, exhibits glassy characteristics at low temperatures \cite{Bako2007},
similar to two-level systems. Our glassy interpretation is
consistent with the reported drop in the oscillator period and
concomitant peak in the dissipation. The microscopic nature of the proposed
glassy state in solid $^4$He is still elusive and outside the scope of
our phenomenological theories for torsional oscillator and specific heat data. 
Although lacking a microscopic model, we can outline some
general properties attributable to this state. We
propose that there is a structural component of the crystal (e.g.,
dislocations, grain boundaries or dislocation cores, etc.) that
has glass-like features and exhibits a freeze-out of dynamics at lowest
temperatures. This glassy component comprises a small fraction of
the sample with characteristic time dynamics.
In addition strong coupling between a glassy and possible supersolid component might be
present in these crystals. For that one would need more detailed experiments
exploring the crystal structure in order to address the physical nature of the
glassy state at lowest temperatures.
Details of the glass model and data analysis can be found in Refs.~\cite{Balatsky2007,Nussinov2007}, 
where it was assumed that the backaction
of the glassy subsystem onto the torsional oscillator is small.

\section{Entropy Dilemma}
The entropy is a state function of a thermodynamic system in equilibrium 
and records the number of excited states. Thus entropy
and specific heat are the preferred quantities for characterizing bulk phase transitions.
Assuming a noninteracting Bose-Einstein condensate (BEC) in three dimensions with a parabolic band, the specific 
heat is 
$C_{BEC}(T) = \frac{15 \zeta(\frac{5}{2})}{4 \zeta(\frac{3}{2})} R\, (T/T_c)^{3/2}$ 
for $T \leq T_c$.
Correspondingly, the entropy at $T_c$ is universally
$S_{BEC}(T_c) = \int_0^{T_c} dT C_{BEC}/T = 
\frac{5 \zeta(\frac{5}{2})}{2 \zeta(\frac{3}{2})} R\, (T/T_c)^{3/2}\simeq 5/4\, R$, 
with gas constant $R=8.314$ J/(mol\ K) and Riemann's $\zeta$ function. 
For comparison, even the entropy of 
strongly correlated superfluid $^4$He at the $\lambda$ point ($T_\lambda=1.8$ K and $P_\lambda=26$ bar) 
is of the same order, namely $S_\lambda \approx 0.55\, R$.
On the other side, a glass comprised of a distribution of two-level systems (TLS)
has a low temperature specific heat
$C_{TLS}(T) \simeq \nu R (T/T_c)$ for $T < T_c$, where $\nu$ is proportional to the  fraction of the TLS in the
sample.  Since only a small fraction of the sample undergoes a 
transition into a BEC or glass phase, we expect that the measured
low temperature specific heat for $T<T_c$ is either of the form
$C = A_{BEC} T^{3/2} + B_{Debye} T^3$ or
$C = A_{glass} T + B_{Debye} T^3$ \cite{Balatsky2007}. 

Figures 1 through 3 show the specific heat recently measured by Lin et al.\ \cite{Lin2008} 
on ultra-pure solid $^4$He contaminated with 1 ppb of $^3$He, which is qualitatively 
consistent with earlier measurements at higher $^3$He concentrations by the PSU group \cite{Lin2008}. 
Indeed, Fig.~1 shows a clear deviation from perfect Debye behavior
below 0.12 K, with the linear-$T$ term visible as a finite intercept in Fig.~2.
Note, that for a BEC the $C/T \to 0$ with $T \to 0$.
Fig.~3 shows the difference $\Delta C/T = (C - C_{Debye})/T$, which is a measure of the entropy. 
We estimate that $S(T_c) \approx  3.3\cdot 10^{-6}\, R$. This value is nearly four orders of magnitude
smaller than is expected for 1\% of solid $^4$He transforming into a supersolid.  Therefore, neither
the low-$T$ behavior of $C$ nor the excess entropy associated with a phase transition are consistent 
with a BEC of a noninteracting or strongly interacting system of vacancies. However, at this point 
we cannot rule out that below $T_c$ a dominant glass and subdominant BEC state coexist.

\bigskip
\begin{figure}[h]
\begin{minipage}[t]{18pc}
\centerline{\includegraphics[width=18pc]{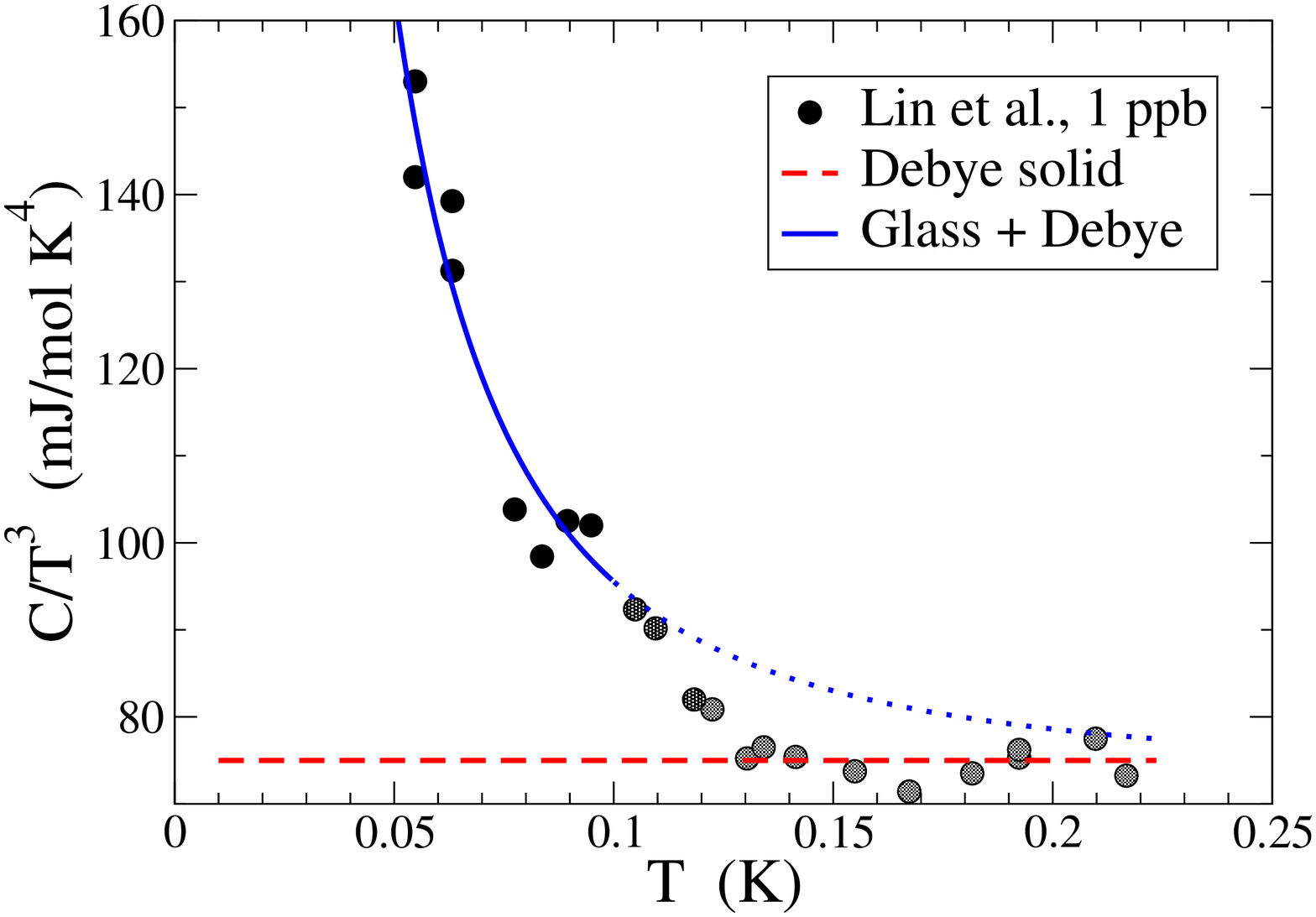}}
\caption{\label{fig1}$C/T^3$ of ultra-pure (1 ppb of $^3$He) solid $^4$He from Lin et al.\ \cite{Lin2008}. 
The non-Debye behavior below $T_c \approx 0.12$ K signals the freezing out of
two-level systems.}
\end{minipage}
\hfill
\begin{minipage}[t]{18pc}
\centerline{\includegraphics[width=17pc]{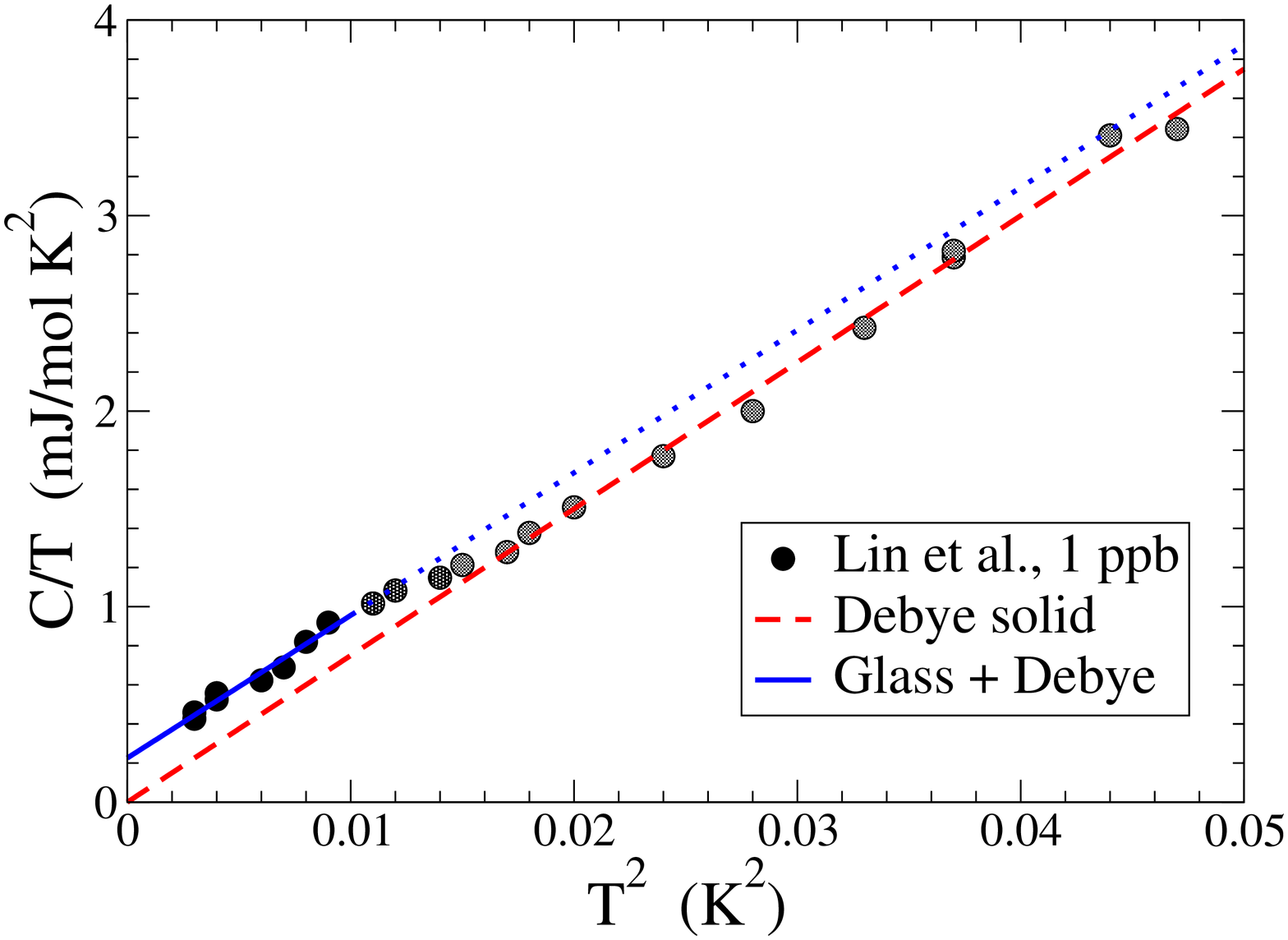}}
\caption{\label{fig2}$C/T$ of ultra-pure solid $^4$He.
Below $T_c$, $C/T$ deviates from the behavior of a perfect Debye solid with a nonzero intercept 
indicative of a small glassy component ($\nu \approx 3.3\cdot 10^{-6}$). }
\end{minipage}
\end{figure}

\begin{figure}[h]
\includegraphics[width=17pc]{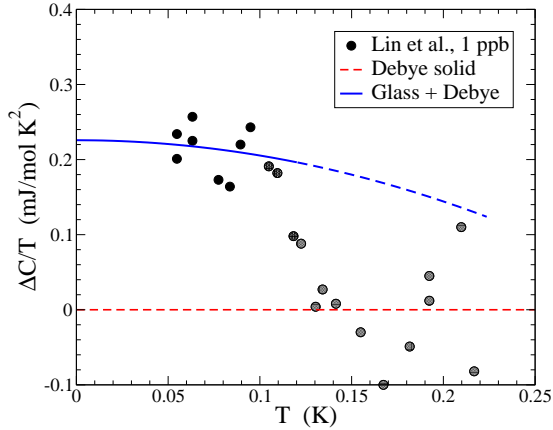}
\hfill
\begin{minipage}[b]{17pc}
\caption{\label{fig3}$\Delta C/T$ of ultra-pure solid 
$^4$He. The estimated entropy associated with the transition
is $S(T_c) \approx 3.3 \cdot 10^{-6}\, R$. This entropy
change is six orders of magnitude less than that for an ideal BEC transition
$S_{BEC}(T_c) \approx 1.25\, R$. Thus it cannot explain the putative supersolid
transition, even when assuming that only 1\% of solid $^4$He transform.}
\end{minipage}
\end{figure}

\section{Torsional Oscillator Response}

A torsional oscillator (TO) measures the susceptibility of the angular response, it does not
directly monitor the moment of inertia of the supersolid. Hence, a change in the response function
(period or dissipation) vs temperature is not sufficient to deduce a change in 
the moment of inertia.
In linear response the angular susceptibility of an underdamped externally driven 
TO in the presence of a small backaction component $g$ is 
$\chi^{-1} = \chi^{-1}_0  - g$ \cite{Nussinov2007}. 
In Fourier space the bare susceptibility of the empty cell is
\begin{equation}
\chi_0^{-1}(\omega; T) = \alpha - i \gamma_{osc}\omega - I_{osc} \omega^2.
\end{equation}
Here $\alpha$ is the restoring coefficient, $\gamma_{osc}$ is the dissipative coefficient and
$I_{osc}$ is the moment of inertia of the TO apparatus. The function $g(\omega; T)$ captures the response 
of added solid $^4$He with a
supersolid or glassy component that acts like a generalized polarization in a dielectric medium.
For a solid with a supersolid subsystem one expects the backaction to be
\begin{equation}
g_{ss}(\omega; T) = i \gamma_{He} \omega + I_{ss}(T) \omega^2, 
\end{equation}
while for a solid with a glassy subsystem it is
\begin{equation}
g_{gl}(\omega; T) = i \gamma_{He} \omega + I_{gl} \omega^2 + g_0( 1 - i\omega\tau)^{-1}.
\end{equation}
The significant difference between both scenarios is that for a supersolid the moment of inertia
$I_{ss}(T)$ is $T$ dependent and accounts for the change in the oscillator's period, while for a glass
$I_{gl}$ is $T$ independent (to leading order) and the $T$ behavior originates from the relaxation
time 
$\tau(T)=\tau_0\exp(\Delta/k_B T)$ in $g_{gl}(\omega; T)$. 
In this expression and in our fits we assumed a simplified Vogel-Fulcher-Tamman relaxation time 
for glasses with an activated behavior, as discussed in more detail in Ref.~\cite{Nussinov2007}. 

In figures 4 and 5 we show the difference between the change in period and dissipation due to
a glass transition vs a supersolid transition in parts of the crystal. 
Comparing our model calculations with TO measurements \cite{Rittner2006} 
we can reproduce both period and dissipation changes if we assume a glassy transition, while this
is not possible for a simple supersolid transition that is due to a change in $I_{He}(T)$ alone.
The glass fit is discussed in detail in Ref.~\cite{Nussinov2007}.  
For the limiting case of a single relaxation mode,
that means a glass exponent of $\beta=1$, we were able to derive the simplified fit functions
for period $P$,
\begin{equation}
P^{-1}(T) = f_0 - B f_0^{-1} [ 1 + (2\pi f_0 \tau(T))^2 ]^{-1},
\end{equation}
and dissipation (inverse of the quality factor $Q$),
\begin{equation}
Q^{-1}(T) = A \tau(T) [ 1 + (2\pi f_0 \tau(T))^2 ]^{-1} + Q_{\infty}^{-1}, 
\end{equation}
with frequency $f_0 = 1/P(0)$.
In Ref.~\cite{Nussinov2007} 
we assumed that fit parameters $A$ and $B$ are independent, however, they obey the constraint
$A/B \simeq 4\pi/ f_0$ \cite{Dorsey2008, Yoo2008}. 
This leads to slightly degraded fits of both period and dissipation compared 
to the unconstrained fits in Ref.~\cite{Nussinov2007}, 
though still in fair agreement with experiment.
A key consequence for a glass is that the dissipation change, 
$\Delta Q^{-1} = Q^{-1}-Q^{-1}_\infty$, is given by the frequency change,
$\Delta P^{-1} = |P^{-1} - P(0)^{-1}|$, according to
$\Delta Q^{-1} \sim 4\pi \tau\ \Delta P^{-1}$, that means, that the maximum change is
$\Delta Q^{-1} \sim 2 \Delta P/P$. 
This result is equivalent to Ref.~\cite{Yoo2008} and similar to a phenomenological model 
by Huse and Khandker, up to a factor of two,
who obtained $\Delta Q^{-1} \sim \Delta P/P$ \cite{Huse2007}.

On the other hand, for the supersolid fit, we assumed the following functional dependence for the moment of 
inertia to describe the measured drop in period (note that the precise shape and form is irrelevant for the 
ensuing discussion),
\begin{equation}
I(T) = I_0 ( 1 - 0.5\epsilon\,(1-\tanh[(T-0.165 \,{\rm K})/0.04\, {\rm K}]) ).
\end{equation}
In order to fit the period change measured by Rittner and Reppy \cite{Rittner2006} 
we assumed that a small fraction $\epsilon \simeq 3.6\cdot 10^{-5}$ of the entire oscillator
moment of inertia,  $I=I_{osc}+I_{He}$, transforms into a supersolid.
This is consistent with $0.1-1\%$ of nonclassical rotational inertia fraction (NCRIF). 
Fig.~5 demonstrates that the
putative supersolid transition seen in Fig.~4 does not account for the simultaneously reported
peak in the oscillator's dissipation. It would require a highly unrealistic temperature
dependence of the dissipative coefficient $\gamma(T) \propto Q^{-1}(T)$.
We used equations (11) and (16) of Ref.~\cite{Nussinov2007} 
to calculate the period $P(T)$ and quality factor $Q(T)$ of
the supersolid with
\begin{equation}
Q_{ss}(T) = \frac{\alpha P(T)}{\sqrt{8}\pi\gamma} \sqrt{ 1+\sqrt{1-(P_c/P(T))^2}},
\end{equation}
and $P_c = (2\pi\gamma/\alpha)^2 < P(T)$.
It follows that to leading order the maximum dissipation of a uniform supersolid is given by
$\Delta Q_{ss}/Q_{ss} \sim \Delta P/P$ or equivalently
$\Delta Q_{ss}^{-1}/Q_{ss}^{-1} \sim \Delta P/P \sim 10^{-5}$.
Thus for a supersolid one expects only a very small change in dissipation $Q_{ss}^{-1}$, as seen in Fig.~5.
This is in clear contrast to the glassy scenario where the maximum change is of order unity, 
$\Delta Q/Q \sim 2 Q_\infty \Delta P/P \sim 1$. 

\bigskip
\begin{figure}[ht]
\begin{minipage}[t]{18pc}
\centerline{\includegraphics[width=18pc]{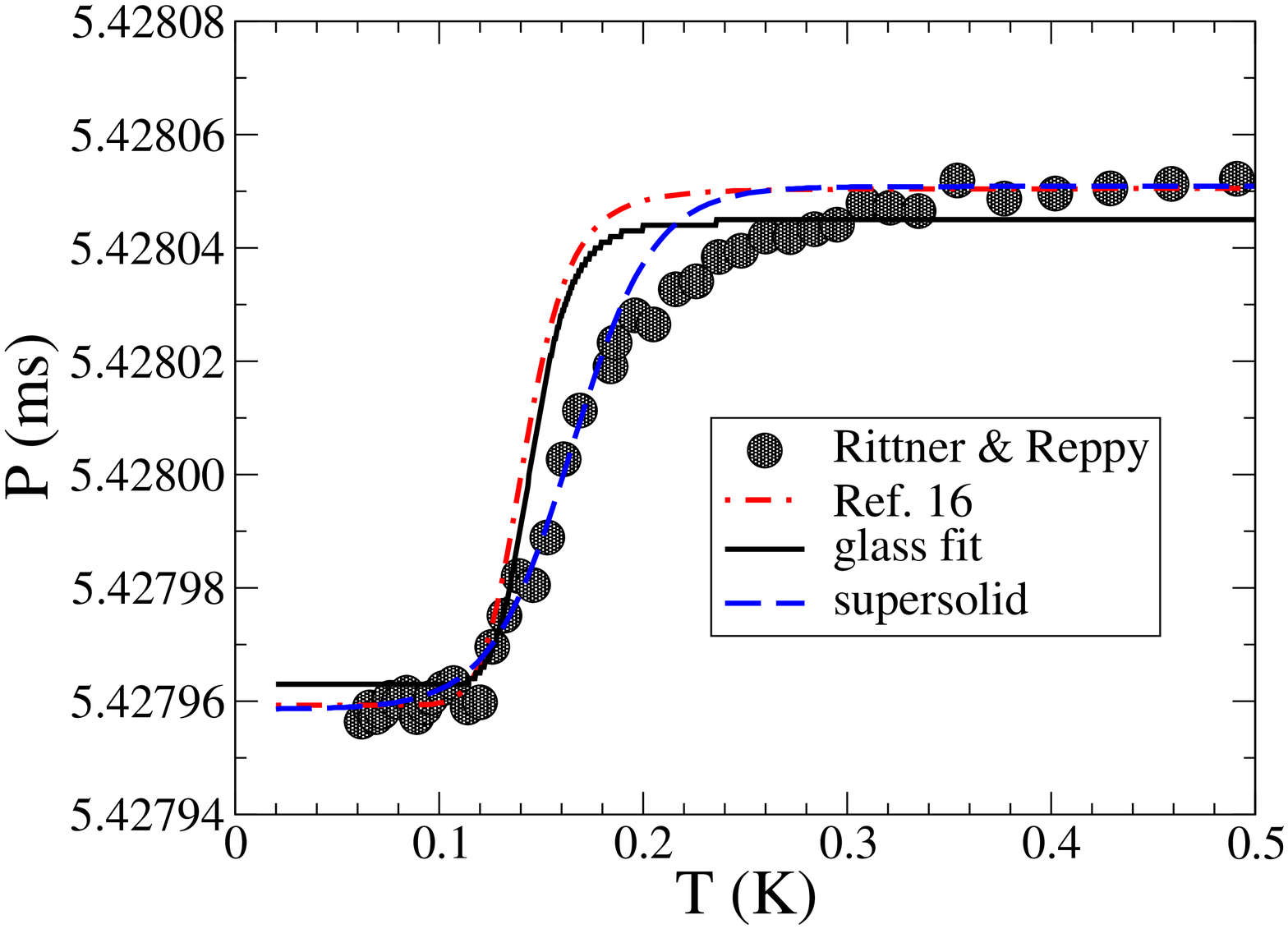}}
\caption{\label{fig4} Change in resonance period $P$ \cite{Rittner2006}. 
Both glass and supersolid model can explain the observed change.
Fit parameters for glass are 
$A=0.0390 \, {\rm s}^{-1}$,
$f_0 = 184.23119$~Hz,
$Q_{\infty}^{-1} = 11.33 \cdot 10^{-6}$,
$\tau_0=0.360 \, \mu$s,
$\Delta=1.135$~K.
}
\end{minipage}
\hfill
\begin{minipage}[t]{18pc}
\centerline{\includegraphics[width=18pc]{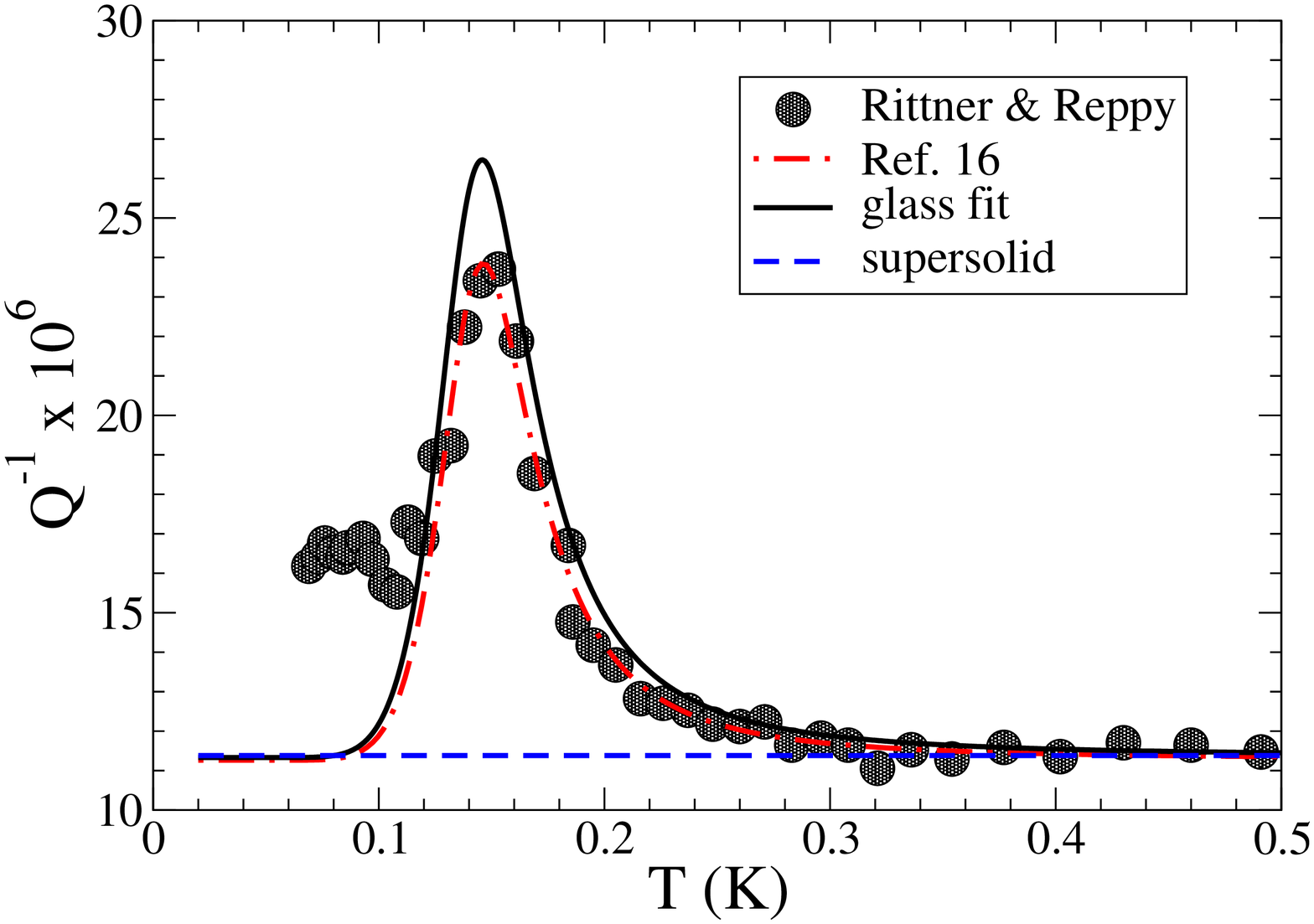}}
\caption{\label{fig5} Change in dissipation (inverse quality factor $Q$) \cite{Rittner2006}. 
Only the glass model can explain the pronounced dissipation peak. Note that for supersolid scenario
the small change in dissipation is not visible on this scale.}
\end{minipage}
\end{figure}

\section{Conclusions}
Our analysis of both the specific heat and the TO measurements for
putative supersolid $^4$He reveals the existence of a dilemma
between the amount of the excess entropy and dynamic response of TO
experiments. We argue that this dilemma can be overcome by invoking
a glass-like transition below $\sim 0.12$ K in ultra-pure solid
$^4$He. Our phenomenological treatment of the specific heat and TO
response does not allow us to identify the microscopic nature of
this anomaly, but we speculate that parts of the solid freeze out
and undergo a glass rather than a BEC transition. This glassy
component may coexist with a possible supersolid component, which
may or may not have been present in the samples analyzed.  Thus it is
reasonable to assume that dislocations give rise to two-level
systems and form a glassy state, which dominantly contribute to the
observed features. At this point, more dynamic studies are needed to
resolve the physical nature of the glass-like phase and if a glass
and supersolid state can coexist in the low temperature phase.

\section{Acknowledgments}
We wish to thank Rittner, Reppy, Aoki, Kojima, Davis, and Goodkind for stimulating discussions
and sharing their unpublished work with us. We are especially grateful to Dorsey for bringing our
attention to the fitting constraint for period and dissipation.
This work was supported by U.S.\ DOE under Contract No.\ DE-AC52-06NA25396. We also thank the
CMI of WU for partial support.


\end{document}